\documentstyle[11pt,aussois2003,epsfig,twoside]{article}
\def\edcomment#1{\iffalse\marginpar{\raggedright\sl#1\/}\else\relax\fi}
\reversemarginpar
\begin{document}
%
\title{Galaxy-galaxy lensing results from COMBO-17}
%
\author{Martina Kleinheinrich, Peter Schneider, Thomas Erben, Mischa Schirmer}
\affil{Institut f\"ur Astrophysik und Extraterrestrische Forschung, 
       Universit\"at Bonn, Auf dem H\"ugel 71, D-53121 Bonn, Germany}

\author{Hans-Walter Rix, Klaus Meisenheimer}
\affil{Max-Planck-Institut f\"ur Astronomie, K\"onigstuhl 17, D-69117 
       Heidelberg, Germany}
\author{Christian Wolf}
\affil{University of Oxford, Astrophysics, Keble Road, Oxford, OX1 3RH,
       UK}
\label{page:first}
\begin{abstract}
 We use galaxy-galaxy lensing to study the statistical properties of the dark
 matter halos of galaxies. Our data is taken from the COMBO-17 survey which
 has imaged 1 square degree in 17 optical filters giving accurate photometric
 redshifts and spectral classification down to $R=24$. This allows us to
 study lens galaxies at  $z=0.2-0.7$. Shapes are measured from deep $R$-band
 images taken at the best seeing conditions ($0\farcs 8$ PSF). We model the 
 lens
 galaxies as singular isothermal spheres (SIS). Investigating the dependence
 of the velocity dispersion on the luminosities of the lens galaxies we are
 able to reproduce the Tully-Fisher/Faber-Jackson relation. Further we find
 a larger lensing signal for early-type galaxies than for the late-types.
 However, when testing the dependence of the velocity dispersion on radius 
 we see clear deviations from the SIS model. The velocity dispersion is first
 rising and then declining again. This is exactly what is expected for lens
 galaxies that can be modeled by Navarro-Frenk-White (NFW) profiles.
\end{abstract}

\section{Introduction}
Rotation curves of spiral galaxies yield the clearest evidence for the 
existence of extended dark matter halos in galaxies. Numerical simulations of
galaxy formation also predict large dark matter halos and make detailed 
predictions about their density profile and its relation to the total mass of
the halos. However, from observations, only little is known in detail about
dark matter halos.
Observations of the central parts of low-surface brightness galaxies seem to 
conflict with the simulations but are difficult to interpret because the 
effects of the luminous matter have to be taken into account. At large 
distances from the centers the dark matter is dominating so that the 
simulations can be tested more directly. But observations of the outer parts 
of galaxies (at several hundred kpc) are still challenging due to the data 
requirements. We probe these regions with weak gravitational lensing. In 
Section 2 we will briefly summarize what is so far known about the dark 
matter halos of galaxies. In Section 3 the COMBO-17 survey will be presented. 
Section 4 introduces the method of galaxy-galaxy lensing and how we use it for
our measurements which are presented in Section 6. Section 5 gives an overview 
over other galaxy-galaxy lensing measurements and how COMBO-17 compares to
previously used data sets. Finally, in Section 7 we present our conclusion and 
an outlook.

\section{Dark matter halos of galaxies}
The outer parts of dark matter halos of galaxies can only be probed 
statistically with the dynamics of satellite galaxies or with galaxy-galaxy
lensing. While satellite studies have to rely on dynamical equilibrium,
galaxy-galaxy lensing can in principle always be applied. Rotation curves of
galaxies or strong lensing are only available for the innermost parts of 
galaxies, up to a few tenth of kpc. 

Flat rotation curves have led to the singular isothermal sphere (SIS) model of 
galaxies which yields a different density profile than the 
Navarro-Frenk-White (NFW) profile favoured by numerical simulations. Most 
studies so far used the SIS model or a truncated SIS model with a steeper 
decline in density in the outer parts of the halos. Galaxy-galaxy lensing 
studies tried to find an outer scale where the density profile starts to 
deviate from the SIS. It was found that dark matter halos stay isothermal out 
to at least $200\mathrm{h}^{-1}\mathrm{kpc}$ implying that halos of individual 
galaxies might overlap (e.g.\ Brainerd, Blandford, \& Smail 1996; Fischer et 
al. 2000; Hoekstra et al. 2002). So far it has not been tested if the SIS 
model or the NFW profile provides a better fit to the data.

From studies of the luminous parts of galaxies the Tully-Fisher relation
for spiral galaxies and the Faber-Jackson relation for elliptical galaxies
have been established according to which the velocity dispersion of galaxies
scales with its luminosity as $\sigma_v\propto L^\eta$ with $\eta\approx 0.25$.
It is not clear if such relations also hold on larger scales. Zaritsky et al.
(1997) find $\eta\approx0$ from satellite studies while McKay et al. (2001) 
find $\eta\approx0.5$ from galaxy-galaxy lensing. If $\eta=0$ then there is
no relation between (aperture) masses of galaxies and their luminosities while
in the case of $\eta=0.5$ the (aperture) mass of a galaxy is proportional 
to its luminosity.

From galaxy-galaxy lensing it was further found that early-type galaxies have 
larger velocity dispersions than late-type galaxies (McKay et al. 2001). 
However, more investigations are still needed to improve our understanding of 
dark matter halos of galaxies, their extent, total mass and density profile 
and how these depend on luminosity, type or redshift of the galaxies.

\section{The COMBO-17 survey}

COMBO-17 ({\bf C}lassifying {\bf O}bjects by {\bf M}edium-{\bf B}and 
{\bf O}bservations in {\bf 17} filters, Wolf et al. 2001, 2003) is a deep 
multi-colour survey that is particularly well suited for studying 
galaxy-galaxy lensing. All observations were carried out with the Wide Field 
Imager at the MPG/ESO 2.2-m telescope on La Silla, Chile, which has a large 
field-of-view of $34\arcmin\times33\arcmin$. The five survey fields are given 
in Table 1. 
\begin{table}[htb]
\begin{tabular}{llll}
Field   & $\alpha_{\mathrm{J2000}}$ & $\delta_{\mathrm{J2000}}$ &  \\ 
\noalign{\smallskip} \hline \noalign{\smallskip} 
A 901   & $09^{\mathrm{h}} 56^{\mathrm{m}} 17^{\mathrm{s}}$ & $-10^{\circ} 
        01' 25''$ & Abell 901/902\\ 
S 11    & $11^{\mathrm{h}} 42^{\mathrm{m}} 58^{\mathrm{s}}$ & $-01^{\circ} 
        42' 50''$ & random field\\
Chandra & $03^{\mathrm{h}} 32^{\mathrm{m}} 25^{\mathrm{s}}$ & $-27^{\circ} 
        48' 50''$ & Chandra Deep Field South \\ 
SGP     & $00^{\mathrm{h}} 45^{\mathrm{m}} 56^{\mathrm{s}}$ & $-29^{\circ} 
        35' 15''$ & South Galactic Pole\\ 
FDF     & $01^{\mathrm{h}} 05^{\mathrm{m}} 47^{\mathrm{s}}$ & $-25^{\circ} 
        51' 27''$ & FORS Deep Field (only BVRI)\\
\noalign{\smallskip} \hline
\end{tabular}
\caption{Survey fields of COMBO-17}
\label{table:fields}
\end{table} 
The observations have
been finished whereas data reduction is still ongoing for some of the fields.
Redshifts and spectral classification for objects brighter than $R=24$ is 
derived from observations in UBVRI and 12 optical medium band filters. Shapes
of galaxies are measured from deep $R$-band images with limiting magnitudes
around $R=25.5$ that were taken at the best seeing conditions (below 
$0\farcs8$ PSF). Therefore COMBO-17 provides an unique data set in which 
redshifts are not only available for the lens galaxies but also for the source 
galaxies. Spectral classification of the lens galaxies allows us to 
distinguish lens galaxies of different types.

\section{Galaxy-galaxy lensing}
The aim of galaxy-galaxy lensing is to study the dark matter halos of lens
galaxies. Due to the effect of lensing, the images of background galaxies are
strechted tangentially with respect to the lens galaxies. Figure 1
illustrates this effect for intrinsically round background galaxies. 
\begin{figure}[htp]
  \begin{center}
   \leavevmode
   \begin{minipage}[l]{0.37\textwidth}  
      \includegraphics[width=\hsize]{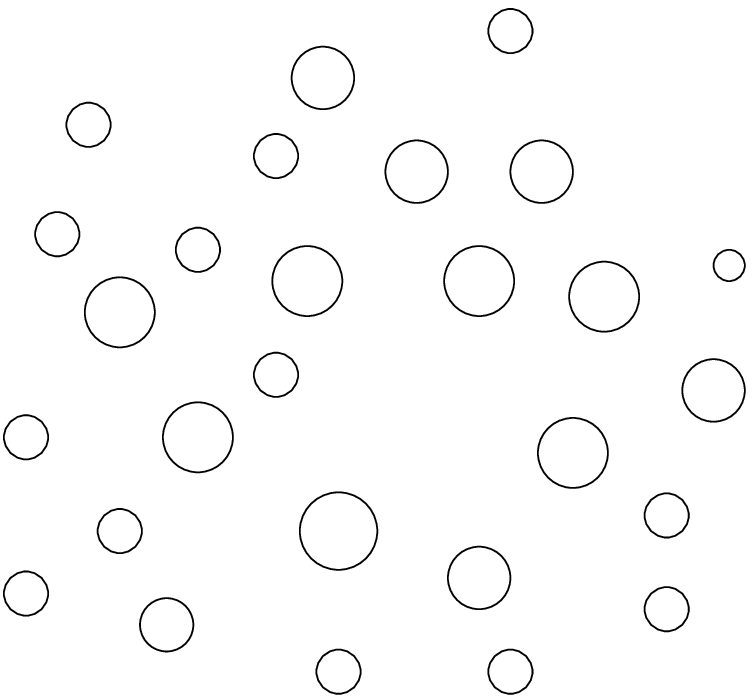}
      \begin{center} {\tiny   without lensing } \end{center} 
      \vspace{0.1mm}
   \end{minipage}
   \hspace*{1cm}  
   \begin{minipage}[r]{0.37\textwidth}
      \includegraphics[width=\hsize]{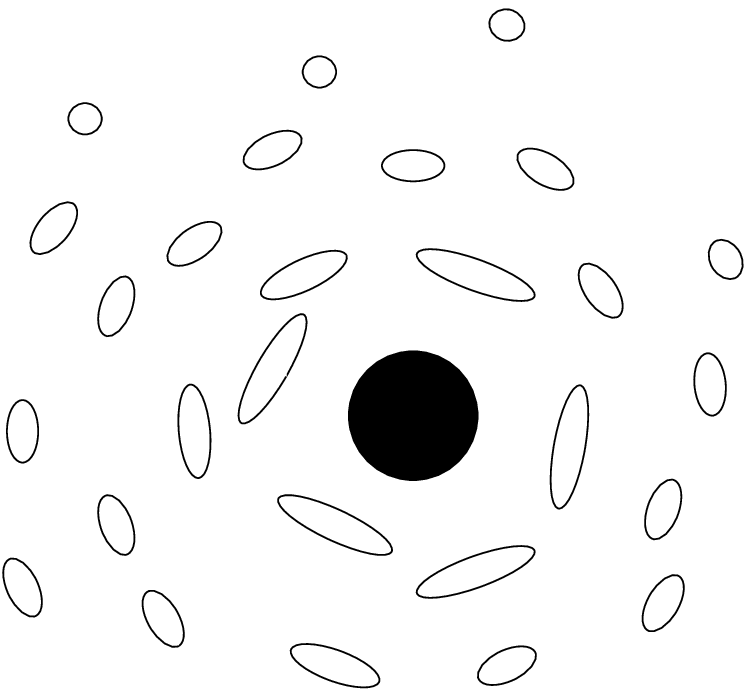}
      \begin{center} {\tiny  with lensing $\Rightarrow$ tangential stretching}
      \end{center} 
      \vspace{0.1mm}
   \end{minipage}
   \end{center}
   \caption{Effect of lensing on intrinsically round background galaxies}
\end{figure} 

The strength of the shear can be calculated for different parametrizations of 
the lens galaxies. As a first step we will use the SIS model. Due to its 
simplicity and because it reproduces the 
flat rotation curves observed in spiral galaxies this model has become quite
popular and has been used in most previous studies of galaxy-galaxy lensing.
The density profile of the SIS is given by
\begin{equation}
   \rho(r)=\frac{\sigma_{v}^{2}}{2\pi G}\frac{1}{r^{2}}~,
\end{equation}
and the shear by
\begin{equation}
   \gamma(\theta)=\frac{2\pi\sigma_{v}^{2}}{c^{2}}\frac{D_{\mathrm{ds}}}{D_{\mathrm{s}}}\frac{1}{\theta}=0.035\left( \frac{\sigma}{155 \mathrm{\tiny kms}^{-1}} \right) ^{2} \left( \frac{\theta}{10''}\right)^{-1}\frac{D_{\mathrm{ds}}}{D_{\mathrm{s}}}~,
\end{equation}
where $\theta$ is the angular separation between lens and source, $\sigma_v$ 
the velocity dispersion and $D_{\mathrm{ds}}$ and $D_{\mathrm{s}}$ are the 
angular diameters distances between lens and source and observer and source, 
respectively. For typical lenses and separations the shear is much smaller 
than the width of the ellipticity distribution of the source galaxies, making 
it impossible to measure the shear for a single lens-source pair. It can only 
be measured by averaging over thousands of pairs and thus over hundreds of 
lens galaxies. Due to the distance factors in Equation (2) the contributions
from different pairs have to be scaled appropiately requiring knowledge of the
redshifts of lens and source galaxies. Further, it has to be considered that
only averaged density profiles of the lens galaxies can be determined. To
account for intrinsic differences in the lens population further knowledge on
the lenses is needed. This then allows one to split the lens sample into 
different subsamples according to e.g.\ luminosity, redshift, spectral type, 
morphology, stellar mass or environment, or to parametrize the dependence of 
the velocity dispersion on these quantities. The Tully-Fisher/Faber-Jackson 
relation is an example of such a parametrization:
\begin{equation}
   \frac{\sigma}{\sigma_{\star}}=\left(\frac{L}{L_{\star}}\right)^{\eta}
\end{equation}
where $L_\star$ is a characteristic luminosity and $\sigma_\star$ the 
velocity dispersion of $L_\star$ galaxies.

\subsection{The Measurement}
The simplest way of measuring galaxy-galaxy lensing is to determine for each 
lens-source pair the alignment of the source with the lens given by the
tangential ellipticity $\epsilon_{t}$ and to measure this alignment as 
function of distance from the center of the lenses. The tangential ellipticity
$\epsilon_t$ corresponds to $\epsilon_1$ that would be measured in a
coordinate frame that is rotated such that the x-axis is tangentially aligned
with respect to the lens. However, often images of background galaxies are
distorted by more than one lens. Therefore, especially at large radii, the
measurement of the density profile of an isolated galaxy is impossible. 

This 
difficulty can be overcome by the maximum-likelihood technique proposed by 
Schneider \& Rix (1997). There the combined shear from all foreground 
galaxies is calculated for a given lens model and different values of its 
parameters. These predictions are then compared to the observed image shapes 
to find the best fitting values of the parameters. However, because all 
foreground galaxies have to be identified, no sources close to the field 
boundaries can be used. So for unfavorable field geometries this method cannot
be applied directly. Advantages are that one can directly account for multiple 
deflections from lenses along the same line-of-sight and from neighbouring 
lenses and that additional sources of shear from e.g.\ clusters can be 
considered easily. We will use this method in our analysis.

\section{COMBO-17 compared to previous data sets used for measurements of 
galaxy-galaxy lensing}
Galaxy-galaxy lensing has been first measured by Brainerd et al. (1996). 
Since then it has been applied to a number of very different data sets ranging
from the very small but deep Hubble Deep Field North (HDFN) to the shallow but 
large Sloan Digital Sky Survey (SDSS). The most detailed analysis that is 
currently available uses the SDSS to systematically study the dependence of 
dark matter halos on luminosity, type and environment of the lens galaxies
(McKay et al. 2001). This is possible from spectroscopy of their lens galaxies 
which is usually not available for deep surveys like COMBO-17. However, the 
SDSS is limited to studies of local galaxies ($z<0.3$). Due to the accurate 
photometric redshifts and the spectral classification, we will be able to 
extend the measurements from SDSS to higher redshifts and to probe the 
evolution of dark matter halos.

COMBO-17 has the further advantage that photometric redshifts are also 
available for all
source galaxies which has so far only been the case for the investigation of
galaxy-galaxy lensing in the HDFN by Hudson et al. (1998).  All other previous 
studies including that on the SDSS data had to use redshift probability 
distributions at least for the source galaxies, some also for the lens 
galaxies. These redshift probability distributions that estimate the redshift 
of a galaxy from its magnitude can be derived from redshift surveys. However, 
for deep data sets the redshift probability distributions have to be 
extrapolated to the faint magnitudes of the source galaxies which could bias 
the results. With COMBO-17 we are not dependent on such extrapolations which 
renders our results more reliable. Further, the knowledge of redshifts will 
decrease the noise because foreground and background galaxies can be better 
separated.

\section{Galaxy-galaxy lensing results from COMBO-17}
To parametrize the lens galaxies we use the SIS model with a Tully-Fisher/
Faber-Jackson relation described by Equations (1), (2) and (3). As 
characteristic luminosity we adopt $L_{\star}=10^{10}L_{\sun}$ where 
$L_{\sun}$ is derived from the absolute magnitude in the SDSS r-band. 
We take all galaxies with $z_{\mathrm{d}}=0.2-0.7$
and $R=18-22$ as lens galaxies but distinguish also between an early- 
and late-type lens sample. This distinction is purely based on the colours
of the galaxies, no morphological information has been used. Source galaxies
are all resolved galaxies (half-light radius greater than size of the PSF) 
with $z_{\mathrm{s}}=0.3-1.55$ and $R=18-24$. We further demand
$z_{\mathrm{s}}>z_{\mathrm{d}}+0.1$ and exclude source galaxies that have a
neighbour within 10 pixel (corresponding to $2.38''$) as their shape 
measurements might be influenced by the neighbour. This selection makes it 
possible that a galaxy is treated as both lens and source, depending on the 
second galaxy in a lens-source pair. $z=1.55$ is the maximum redshift that 
can be reliably assigned in the classification given the template libraries 
and the filter set of COMBO-17. We will use only two of the survey fields: 
the A 901 field and the S 11 field.

First we fit $\sigma_\star$ and $\eta$ simultaneously. We tried different
separations and finally ended up using all pairs with 
$20\mathrm{h}^{-1}\mathrm{kpc}$$<r<150\mathrm{h}^{-1}\mathrm{kpc}$ which 
yields the most significant results. Figure 2 shows the likelihood contours
for all lenses and for early-type and late-type galaxies separately. The 
contours correspond to 1-$\sigma$, 2-$\sigma$ and 3-$\sigma$. They have been
determined directly from the likelihoods of the different values for the
parameters.
\begin{figure}[htb]
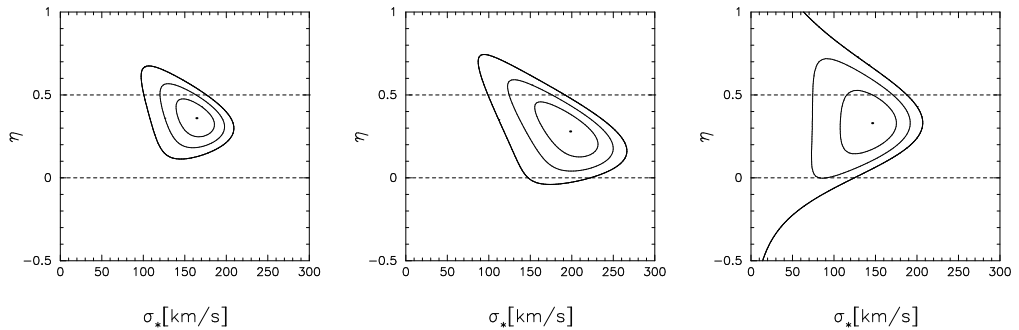

  \begin{center}
   \leavevmode
      \includegraphics[angle=-90,width=0.31\hsize]{figure3.ps}
\hspace*{2mm}
      \includegraphics[angle=-90,width=0.31\hsize]{figure4.ps}
\hspace*{2mm}
      \includegraphics[angle=-90,width=0.31\hsize]{figure5.ps}
   \end{center}
\caption{Fitting $\sigma_\star$ and $\eta$ simultaneously for all lens galaxies
(left panel), early-type lenses (middle panel) and late-type lenses (right 
panel). The contours correspond to 1-$\sigma$, 2-$\sigma$ and 3-$\sigma$.}
\end{figure} 
The best-fitting parameters together with their 1-$\sigma$ limits are
\renewcommand{\arraystretch}{1.5}
\begin{equation}
  \begin{array}[c]{lll}
  \sigma_{\star} = 164^{+20}_{-24} \mathrm{~km~s}^{-1}, & 
     \quad \eta  =  0.36^{+0.11}_{-0.12} & \quad \textrm{all lenses}\\
  \sigma_{\star} =  198^{+32}_{-42} \mathrm{~km~s}^{-1}, & 
     \quad \eta  =  0.28^{+0.17}_{-0.16} & \quad \textrm{early-type lenses}\\
  \sigma_{\star}  =  146^{+28}_{-38} \mathrm{~km~s}^{-1}, & 
     \quad \eta  =  0.33^{+0.18}_{-0.16} & \quad \textrm{late-type lenses}\\
  \end{array}
\end{equation}
\renewcommand{\arraystretch}{1.0}
This shows that early-type galaxies have larger velocity dispersions while no
clear dependence of $\eta$ on galaxy type is seen.

When trying different ranges of allowed separations between lens and source 
galaxies we found that the best-fit velocity dispersion becomes smaller when 
inluding pairs with larger separations. This already indicates that the 
velocity dispersion of the dark matter halos is not constant but decreasing.
We tested this assumption by
fitting the velocity dispersion for separations 
$20\mathrm{h}^{-1}\mathrm{kpc}$$<r<r_{\mathrm{max}}$ where we varied only 
$r_{\mathrm{max}}$ but fixed $\eta=0.35$. Figure 3 shows the results, again for
all lens galaxies, early-types and late-types.
\begin{figure}[hbt]
  \begin{center}
   \leavevmode
      \includegraphics[width=0.31\hsize]{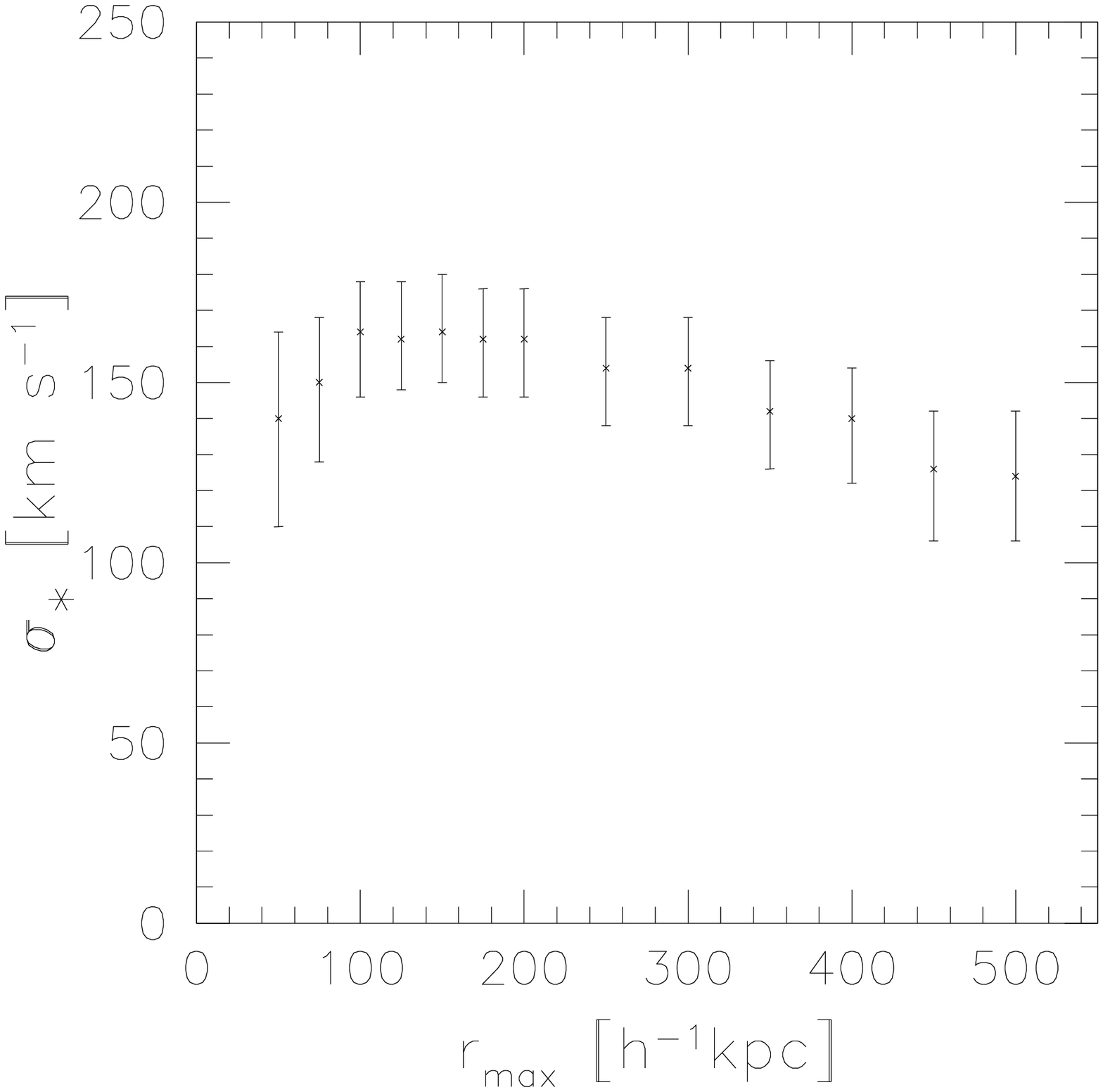}
\hspace*{2mm}						    
      \includegraphics[width=0.31\hsize]{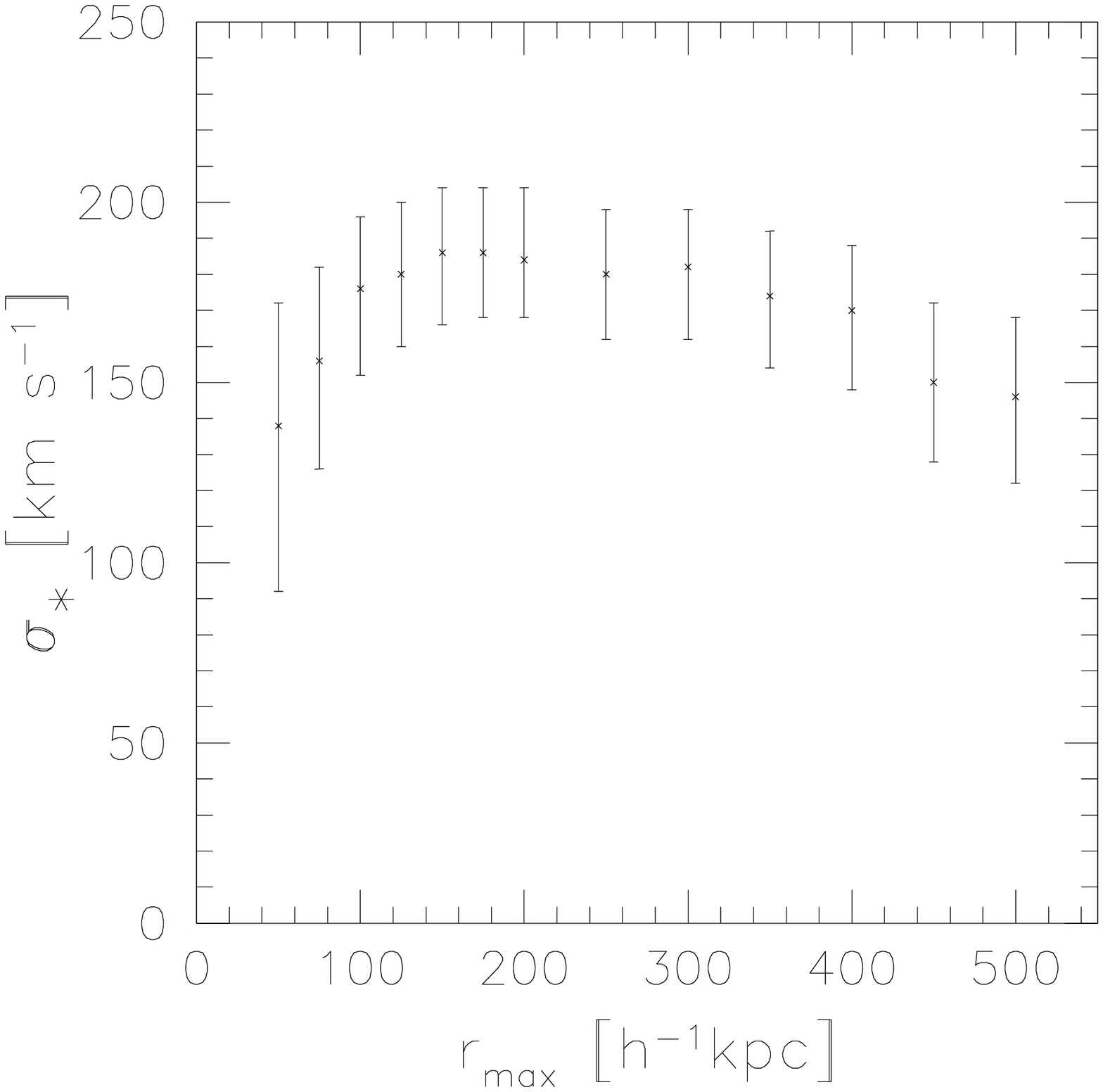}
\hspace*{2mm}						    
      \includegraphics[width=0.31\hsize]{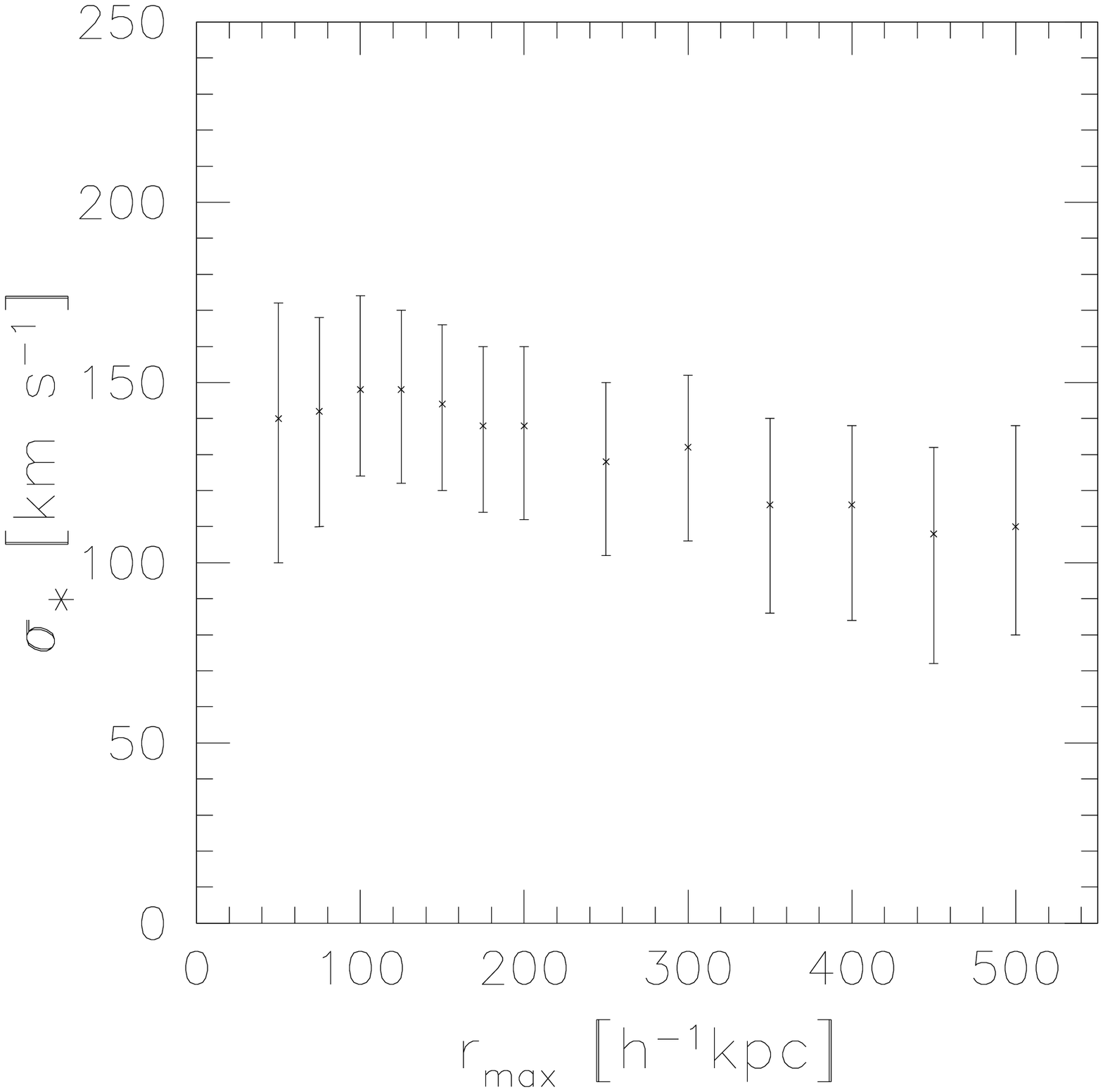}
   \end{center}
\caption{Fitting $\sigma_\star$ with fixed $\eta=0.35$ for different 
separations of lens and source galaxies. For each data point only pairs with 
$20\mathrm{h}^{-1}\mathrm{kpc}$$<r<r_{\mathrm{max}}$ are used. The left panel 
shows the results for all lens galaxies, the middle panel for only early-type 
lenses and the left panel for late-type lens galaxies. The errorbars are 
1-$\sigma$ limits.}
\end{figure} 
Figure 3 clearly shows that the SIS is not a good model at all. Outside about
$r=150\mathrm{h}^{-1}\mathrm{kpc}$ the best-fit velocity dispersion is 
continously falling and is larger for the early-type sample. But probably most 
interesting is the rise in velocity dispersion in the inner parts of the halos
which is most clearly seen for the early-type sample. A velocity dispersion
that is first rising and then declining is exactly what is expected for dark 
matter halos that can be described by NFW halos!

\section{Conclusions and Outlook}
Using the most simple model for describing dark matter halos of galaxies,
the SIS model, we have measured the velocity dispersions of galaxies with 
galaxy-galaxy lensing and tested how these depend on luminosity and spectral 
type.
We find $\sigma_\star\approx 160~\mathrm{km~s}^{-1}$ and $\eta\approx0.35$ 
averaged over all lens galaxies and larger velocity dispersions for early-type
galaxies. These results are in good agreement with previous findings.

However, we also find that the SIS is not a good model for describing dark 
matter halos of galaxies but that the NFW profile might be more appropriate.
Therefore our results on the mean velocity dispersion or the relation between 
velocity dispersion and luminosity or spectral type of the lens galaxies can 
only be regarded as preliminary because these results might depend 
sensitively on the range in separation over which the fit is performed. We will
test if the NFW profile provides a better fit to the data. Having found a 
better parametrization of the radial profile of dark matter halos we can then
calculate aperture masses and reanalyse the dependencies of the density profile
or mass on luminosity and spectral type of the galaxies.

Once the whole data set from COMBO-17 is available we can improve the
statistics and also test the dependence of halo properties on environment,
stellar mass or redshift of the galaxies.

\acknowledgements
This work was supported by the BMBF/DLR project 50 OR 0106, the BMBF/DESY 
project O5AE2PDA/8, and the DFG under the project SCHN 342/3-1.

\label{page:last}
\end{document}